\documentclass[12pt]{article}
\usepackage{amssymb}
\usepackage{latexsym}
\usepackage{amsmath}
\usepackage{graphicx}

\usepackage{esdiff}
\usepackage{mathtools}

\def\o{\omega}
\def\a{\alpha}
\def\ve{\varepsilon}
\def\g{\gamma}
\def\b{\bar{\beta}}
\title{\LARGE{$1+1$-dimensional Dirac oscillator with deformed algebra with minimal uncertainty in position and maximal in momentum }}
\author{M. M. Stetsko\footnote{E-mail: mstetsko@gmail.com}\
\\
  {\small Department for Theoretical Physics, Ivan Franko National University of Lviv,}\\
{\small 12 Drahomanov Str., Lviv, UA-79005, Ukraine
         }}

\begin{document}
\maketitle

\abstract{$1+1$-dimensional Dirac oscillator with minimal uncertainty in position and maximal in momentum is investigated. To obtain energy spectrum SUSY QM technique is applied. It is shown that the Dirac oscillator has two branches of spectrum, the first one gives the standard spectrum of the Dirac oscillator when the parameter of deformation goes to zero and the second branch does not have nondeformed limit. Maximal momentum brings an upper bound for the energy and it gives rise to the conclusion that the energy spectrum contains a finite number of eigenvalues. We also calculate partition function for the spectrum of the first type. The partition function allows us to derive thermodynamic functions of the oscillator which are obtained numerically.}

\section{Introduction}
Dirac oscillator, originally discovered in the paper \cite{Ito_NC67} and rediscovered after two decades \cite{Moshinski_JPA89} is an object of intensive examination during recent two decades. This active interest has its roots in two main reasons. The first one is caused by the fact that the Dirac oscillator is an example of relativistic exactly solvable problems and  different aspects of this problem can be well-studied \cite{Moreno_JPA89,Bentez_PRL90,Villalba_PRD94,Moshinski_JPA95,Szmytkowski_JPA01}. The second reason which has become important for the last decade is motivated by proposals of experimental engineering of this system \cite{Bermudez_PRA07,Longhi_Opt10} which has been finally realized, but just in the case of two dimensional Dirac oscillator \cite{Franco_PRL13}. 

Quantum mechanical problems with deformed commutation relations is an area of active research for recent years. It is supposed that deformed commutation relations can take into account some underlying effects of quantum gravity effectively and thus it is interesting to find out what is the influence of quantum gravitational effects on standard problems of quantum mechanics. Up to date there is no preferable type of deformations of commutations relations, any sort of deformation can take into consideration some specific features of underlying theory. One of the most interesting kinds of deformation was proposed by Kempf and colleagues \cite{Kempf_PRD95} which introduces minimal length into quantum mechanics. From the other side generalization of standard special relativity which also implement effects of quantum gravity gives rise to maximal momentum \cite{Magueijo_PRL02}. Incorporation of the minimal length as well as of the maximal momentum into quantum mechanics leads to corresponding deformed algebra  which was proposed almost a decade ago \cite{Ali_PLB09}. This algebra is also a subject of active investigation and was applied to some problems in quantum mechanics, relativity and cosmology \cite{Nozari_PRD12,Chemissany_JCAP11,Majumder_PRD11}. It should be pointed out that the Dirac oscillators in deformed theories with minimal length (as well as with minimal momentum) have been solved exactly \cite{Quesne_JPA05,Quesne_JPA06,Stetsko_JMP15,Valtancoli_JMP17} and some of its thermal properties have been studied \cite{Menculini_PRD15,Boumali_arx15}.

In our work we investigate $1+1$-dimensional Dirac oscillator in the theory with minimal uncertainty in position as well as maximal uncertainty in momentum. This paper is organized as follows. In the second section we derive energy spectrum of the Dirac oscillator. In the third section we calculate partition function and investigate some thermal properties of the oscillator. Finally, the forth section contains some conclusions. 
\section{Dirac oscillator Hamiltonian and its eigenvalues}

In this work $1+1$--dimensional Dirac oscillator will be considered and corresponding eigenvalue problem will be solved. The eigenvalue equation of the $1+1$--dimensional Dirac oscillator takes the form:
\begin{equation}\label{DO_eq}
H_{DO}\psi=(\sigma_{x}(p_{x}-im\o\beta x)+m\beta)\psi=E\psi
\end{equation}
and here $m$, $\o$ are the mass and frequency of oscillator respectively, $\sigma_x$ and $\beta=\sigma_z$ are the Pauli matrices. We point out that in this work the system of units where $c=\hbar=1$ is used. Operators of position and momentum satisfy deformed commutation relations of the form:
\begin{equation}\label{algebra}
[x,p_x]=i(1-\a p_x+2\a^2p^2_x)
\end{equation}
where $\a$ is a parameter of deformation. It  was shown that the deformed algebra gives rise to minimal uncertainty in position an maximal uncertainty in momentum \cite{Ali_PLB09,Nozari_PRD12}.

The eigenfunction $\psi$ takes the form of a two-component spinor:
\begin{equation}\label{wave_funct}
\psi=\begin{pmatrix}
\psi_1 \\
\psi_2 \\
\end{pmatrix}
\end{equation}
Having used the latter relation for the eigenspinor we can represent the eigenvalue equation (\ref{DO_eq}) in the following form:
\begin{eqnarray}\label{coupled_eq}
(p_x+im\o x)\psi_2=(E-m)\psi_1,\\\label{coupled_eq2}(p_x-im\o x)\psi_1=(E+m)\psi_2.
\end{eqnarray}
The given above system can be decoupled easily, namely for the upper component we derive:
\begin{equation}\label{dec_eq_1}
(p_x+im\o x)(p_x-im\o x)\psi_1=(E^2-m^2)\psi_1.
\end{equation}
To obtain spectrum and eigenfunctions of Dirac oscillator one should use some representation of the deformed algebra. Here we use the following representation:
\begin{equation}\label{rep}
p_x=p, \quad x=i(1-\a p+2\a^2p^2)\frac{\partial}{\partial p}.
\end{equation}
We note that the operators of position and momentum should be hermitian (symmetric) and to provide the hermicity one should use the scalar product with a weight function, namely it takes the form:
\begin{equation}\label{inn_prod}
\langle\psi|\varphi\rangle=\int^{+P_{b}}_{-P_{b}}\frac{dp}{(1-\a p+2\a^2p^2)}\psi^{*}(p)\varphi(p)
\end{equation} 
and here $P_{b}$ denotes boundary value of the momentum which is supposed to be present due to the existence of the maximal uncertainty in momentum. It should be pointed out that this boundary value was assumed to be Planck's momentum \cite{Nozari_PRD12}.

To obtain energy spectrum of the Dirac oscillator we apply SUSY QM technique \cite{Cooper_PR95}. Here we follow the general scheme of SUSY QM and construct a pair of operators $b^+$ and $b^-$ playing the role of raising and lowering operators correspondingly. Taking into account the system of equations (\ref{coupled_eq})-(\ref{coupled_eq2}) this pair of operators can be introduced as follows:
\begin{equation}
b^{\pm}=p\pm im\o x
\end{equation}
The decoupled equation (\ref{dec_eq_1}) can be rewritten in terms of the introduced above ladder operators:
\begin{equation}
b^{+}b^{-}\psi_{1}=(E^2-m^2)\psi_1.
\end{equation}
The decoupled equation for the lower component of the wavefunction $\psi_2$ can be written in similar way. 

To obtain energy spectrum of the Dirac oscillator we assume that the ground  state energy satisfies the condition $E^2-m^2=0$ (zero ground state energy). Now we introduce  the operators (members of SUSY QM hierarchy):
\begin{equation}\label{ladd_op}
b^{\pm}(\xi_i,k_i)=\xi_i p+k_i\mp m\o(1-\a\g p+2\a^2p^2)\frac{\partial}{\partial p}
\end{equation}
and here $\xi_i$ and $k_i$ are some iterative parameters and $\g$ is a dimensionless numerical parameter ($0\leqslant\g\leqslant 1$) introduced in order to compare our results with results obtained for deformed algebra with minimal uncertainty in position only ($\g=0$). In final relations for the case of the algebra with minimal uncertainty in position as well as maximal in momentum one should set that $\g=1$.
Having used the latter operators and taking into account shape invariance condition we can write:
\begin{equation}\label{shape_inv}
b^{-}(\xi_i,k_i)b^{+}(\xi_i,k_i)=b^{+}(\xi_{i+1},k_{i+1})b^{-}(\xi_{i+1},k_{i+1})+\ve_{i+1}
\end{equation}
Form the latter equation it follows that the iterative parameters satisfy the conditions:
\begin{eqnarray}\label{it_1}
\xi^2_{i+1}-2m\o\a^2\xi_{i+1}=\xi^2_{i}+2m\o\a^2\xi_{i};\\\label{it_2}2k_{i+1}\xi_{i+1}+m\o\a\g\xi_{i+1}=2k_{i}\xi_{i}-m\o\a\g\xi_{i};\\ \label{it_3}k^2_{i+1}-m\o\xi_{i+1}+\ve_{i+1}=k^2_{i}+m\o\xi_{i}.
\end{eqnarray}  
Having solved the iterative system of equations written above we can write:
\begin{equation}
\xi_i=1+2im\o\a^2; \quad k_i=-\frac{im\o\a\g(1+im\o\a^2)}{1+2im\o\a^2}; \quad \ve_{i+1}=k^2_i-k^2_{i+1}+m\o(\xi_i+\xi_{i+1}).
\end{equation}
The parameter $\ve_{i+1}$ defines the energy gap between $i$-th and $i+1$-th energy levels. Finally, we write the relation for the energy spectrum in the form:
\begin{equation}\label{energy_z}
E^2_n-m^2=\sum^{n}_{i=0}\ve_i=2m\o n(1+m\o\a^2n)-m^2\o^2\a^2\g^2n^2\frac{(1+m\o\a^2 n)^2}{(1+2m\o\a^2 n)^2}
\end{equation}
In the limit $\a\rightarrow 0$ we obtain $E^2_n-m^2=2m\o n$ which is in agreement with the results of standard quantum mechanics and when $\g=0$ the second term in the latter relation disappears and we recover the relation for the energy spectrum in the theory with minimal length. It is worth being noted that the right hand side of the latter relation is a monotonously increasing function of $n$ and it means that the energy spectrum $E_n$ is also monotonously increasing  function of the quantum number $n$. In contrast to the ordinary quantum mechanics and deformed theory with minimal length in our case we have an upper bound for momentum \cite{Nozari_PRD12} which has been used in the written above relation (\ref{inn_prod}) and this fact leads to the consequence that we have corresponding upper bound for the energy (\ref{energy_z}) and as a result we have finite number of eigenstates. The other important point one should check here is normalizability of corresponding eigenfunctions, especially the eigenfunction of the ground state, which can be derived from the relation:
\begin{equation}\label{gr_st_wf}
b^{-}\psi_1=0.
\end{equation}
Using the evident form for the operator $b^{-}$ and solving corresponding differential equation we arrive at:
\begin{equation}
\psi_{1(0)}=C_{1(0)}(1-\a p+2\a^2p^2)^{-1/(4m\o\a^2)}\exp{\left[\frac{-1}{2\sqrt{7}m\o\a^2}\arctan{\left(\frac{4\a p-1}{\sqrt{7}}\right)}\right]}.
\end{equation} 
and here $\psi_{1(0)}$ denotes the upper component of the ground state energy eigenfunction and $C_{1(0)}$ is the corresponding normalization constant. It is easy to persuade oneself that the written above wave function is nonsingular and normalizable and it means that it is acceptable eigenfunction. The relation (\ref{coupled_eq2}) can be used to show that the lower component of the ground state eigenfunction $\psi_{2(n)}$ is also acceptable.

Apart of the considered above case with zero ground state energy ($E^2-m^2=0$) we also consider the other case when the ground state has nonzero energy $E^2-m^2\neq 0$. To examine this instance one should redefine the ladder operators instead of previously used ones $b^{\pm}$. We introduce these new operators as follows:
\begin{equation}
\bar{b}^{\pm}=\xi'p+k'\mp m\o(1-\a p+2\a^2p^2)\frac{\partial}{\partial p},
\end{equation} 
where $\xi'$ and $k'$ are some constant parameters. The introduced above operators $\bar{b}^{\pm}$ should satisfy the following relation:
\begin{equation}
\bar{b}^{+}\bar{b}^{-}+\bar{\ve}=b^+b^-
\end{equation}
The latter relation gives rise to the equations for the parameters $\xi'$ and $k'$ which take the form:
\begin{eqnarray}
\label{xi_pr}{\xi'}^2-2m\o\a^2\xi'=1-2m\o\a^2,\\\label{k_pr} \xi'(2k'+m\o\a)=m\o\a,\\\label{ve_bar} {k'}^2-m\o\xi'+\bar{\ve}=-m\o
\end{eqnarray}
From the relation (\ref{xi_pr}) it follows that:
\begin{equation}
\xi'_1=1, \quad \xi'_2=2m\o\a^2-1.
\end{equation}
It is easy to check that for the first case $\xi'_1=1$ we obtain $k'_1=0$ and as a result we come back to the previously considered situation of zero ground state energy. The second variant $\xi'_2=2m\o\a^2-1$ leads to nontrivial result for the parameter $k'$ which takes the form: $k'=m\o\a(1-m\o\a^2)/(2m\o\a^2-1)$. And finally for the parameter $\bar{\ve}$ we arrive at the relation:
\begin{eqnarray}\label{gr_st_en}
\bar{\ve}=2m\o(m\o\a^2-1)-m^2\o^2\a^2\frac{(1-m\o\a^2)^2}{(2m\o\a^2-1)^2}.
\end{eqnarray}
The written above relation defines the ground state energy. Even in case $\a=0$ the latter relation is nonzero, but as it will be shown later for the standard commutation relations one obtains the energy spectrum unbounded from below, thus we will have unphysical solution. We also impose the following condition on the parameters of the oscillator: $2m\o\a^2>1$ in order to avoid some singularities in the following relation for the energy spectrum. To calculate the energy spectrum we use the same procedure as above. Having used the system of equations (\ref{it_1})-(\ref{it_2}) we obtain:
\begin{equation}
\xi_{i}=2(i+1)m\o\a^2-1,\quad k_{i}=\frac{(i+1)m\o\a\g(1-(i+1)m\o\a^2)}{(2(i+1)m\o\a^2-1)}.
\end{equation} 
The general relation for the parameter $\ve_{i+1}$ is again defined by the relation (\ref{it_3}). Taking into consideration the written above relations and the relation (\ref{gr_st_en}) the expression for the energy spectrum can be represented in the form:
\begin{equation}\label{en_sp_2}
E^2_n-m^2=2m\o(n+1)(m\o\a^2(n+1)-1)-m^2\o^2\a^2\g^2(n+1)^2\frac{(1-m\o\a^2(n+1))^2}{(2(n+1)m\o\a^2-1)^2}.
\end{equation}
It should be pointed out here that when $\g=0$ (when the minimal length exists only) the second term of the latter relation disappears and again the energy spectrum for the Dirac oscillator is recovered. As it was mentioned above when $\a=0$ we have unbounded form below eigenvalues and it means that the spectrum (\ref{en_sp_2}) becomes unphysical and we conclude that that in case of the standard commutation relations the only one branch of the spectrum (\ref{energy_z}) exists, whereas for the deformed case we have two branches of spectrum.  

One can also use the relation (\ref{gr_st_wf}) but now written for the operator $\bar{b}^{-}$ in order to find the eigenfunction of the ground state with nonzero energy (\ref{gr_st_en}). After simple calculations we can write:
\begin{equation}
\bar{\psi}_{1(0)}=\bar{C}_{1(0)}(1-\a p+2\a^2p^2)^{-\frac{2m\o\a^2-1}{4m\o\a^2}}\exp\left[\frac{-1}{2\sqrt{7}m\o\a^2(2m\o\a^2-1)}\arctan\left(\frac{4\a p-1}{\sqrt{7}}\right)\right].
\end{equation}
Similarly to the previously considered case the written above function is normalizable.
\section{Thermal properties of the Dirac oscillator}
\begin{figure}
\centerline{\includegraphics[scale=0.28,clip]{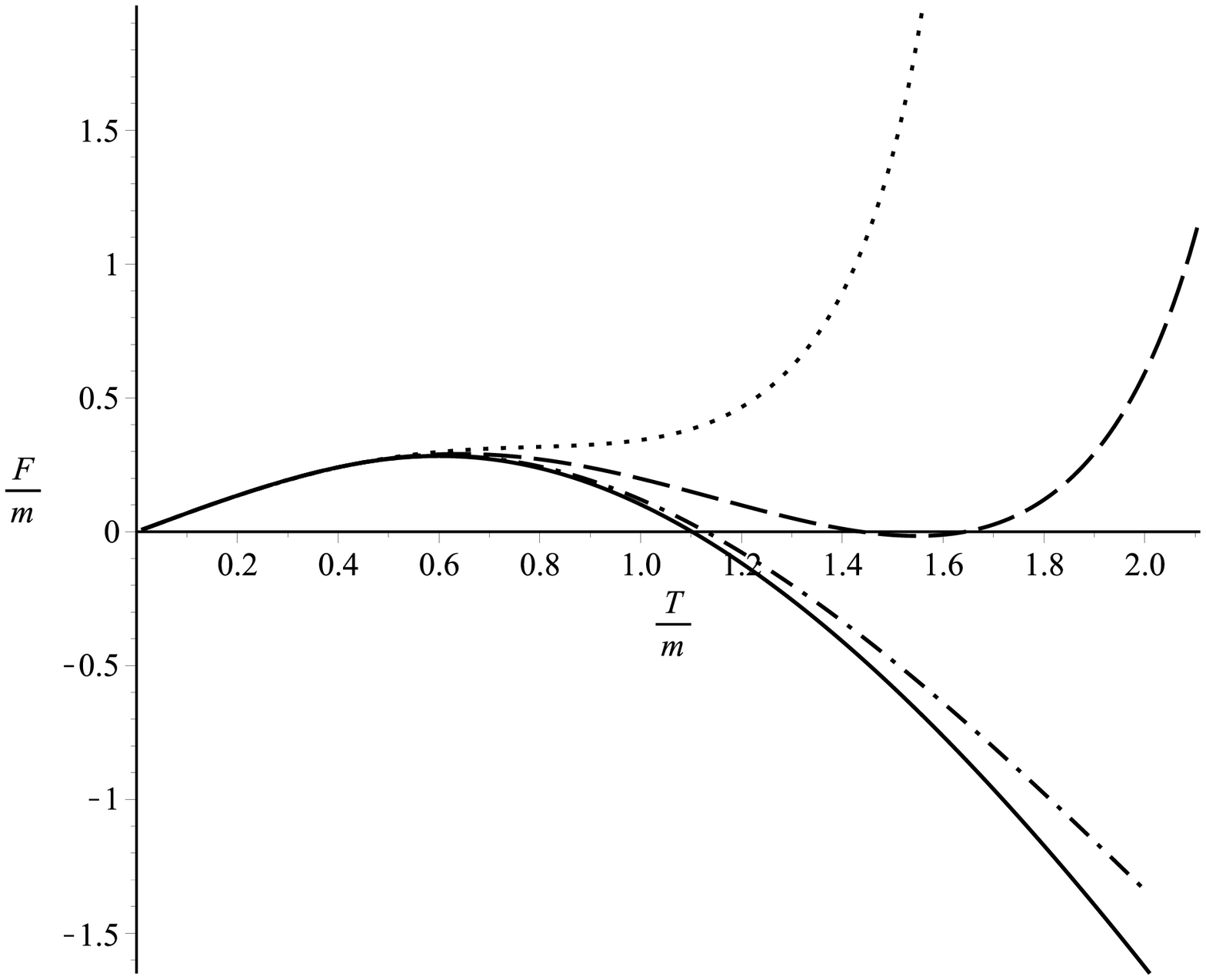}\includegraphics[scale=0.28,clip]{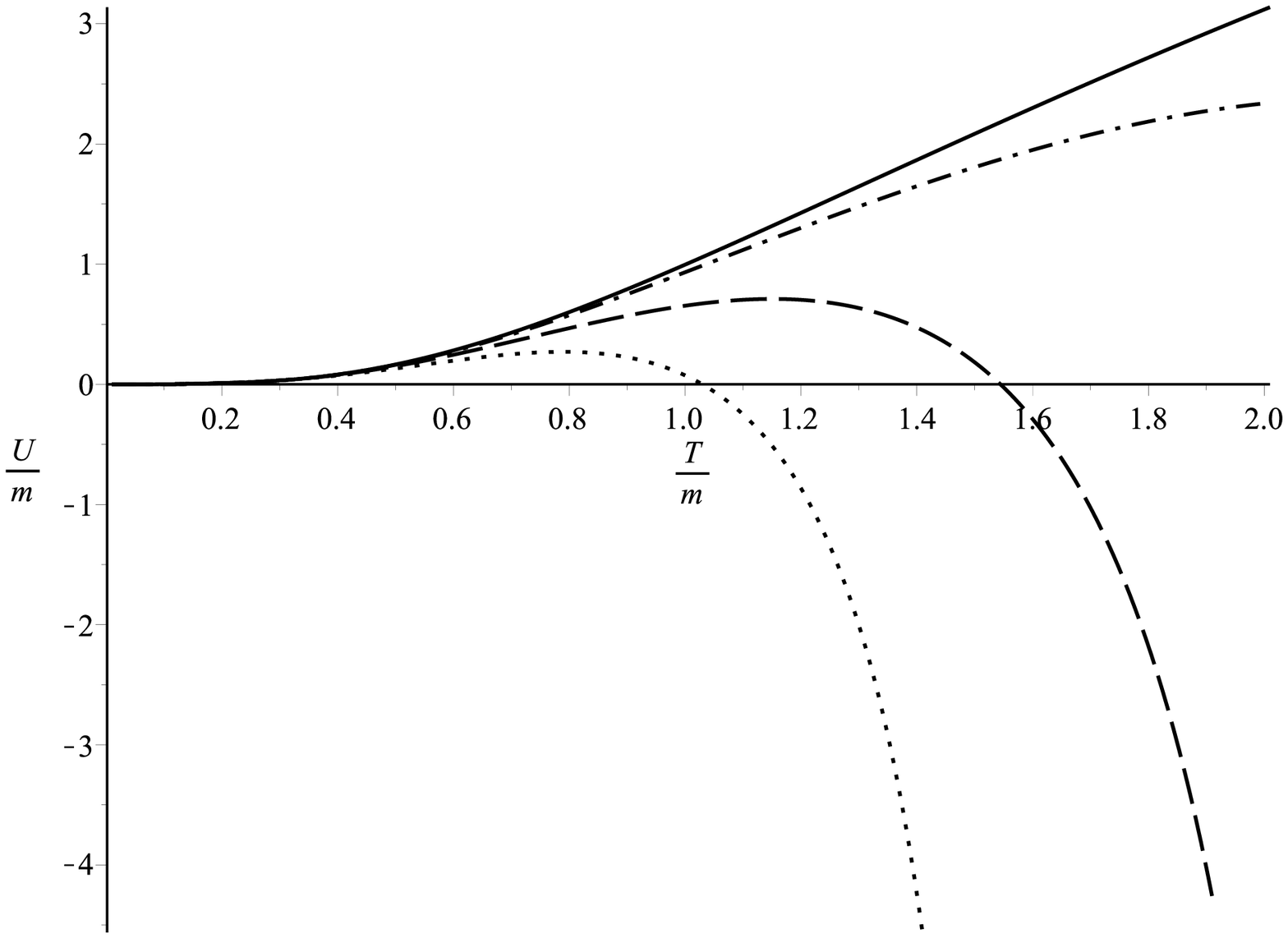}}
\centerline{\includegraphics[scale=0.28,clip]{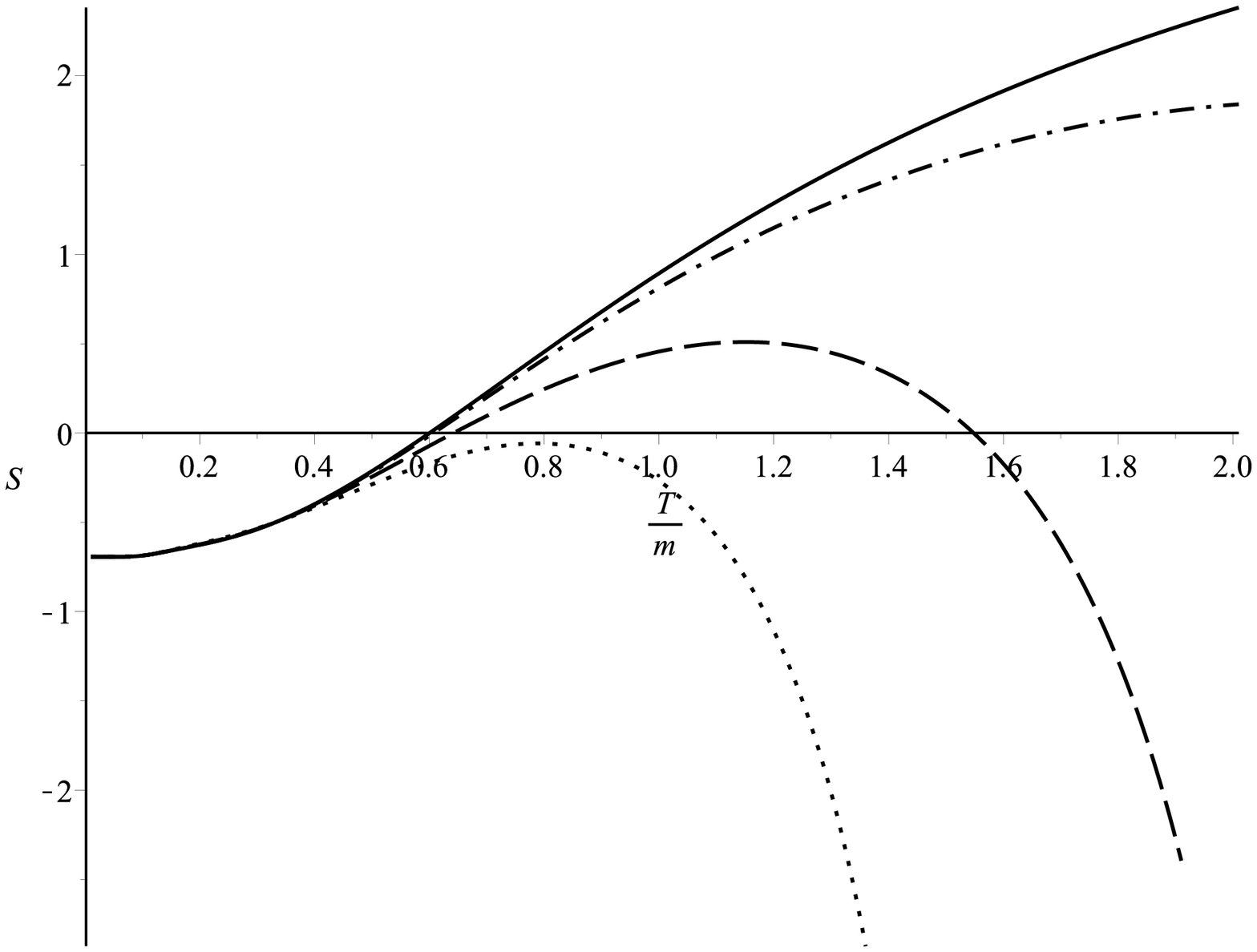}\includegraphics[scale=0.28,clip]{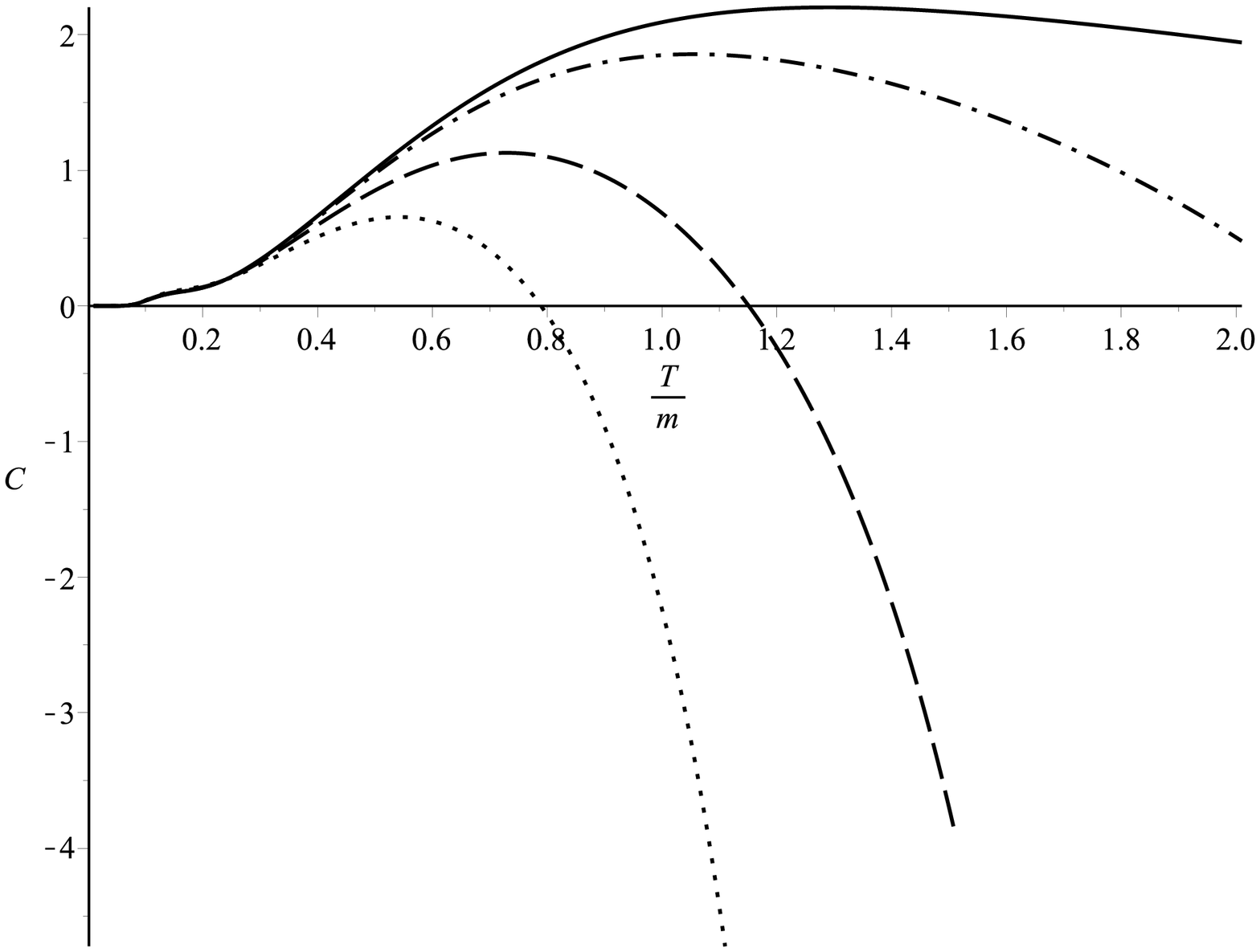}}
\caption{Thermodynamic functions of the Dirac oscillator, namely the free energy per mass $F/m$, internal energy per mass $U/m$, entropy $S$ and heat capacity $C_v$ as the functions of dimensionless temperature $T/m$ for fixed $m$ and $\o$ and several values of the parameter of deformation $\a$. Namely, for all graphs we have taken $m=1$, $\o=2$ and correspondence of the curves is as follows: $\a=0.05$, $\a=0.1$, $\a=0.2$ and $\a=0.3$ are represented by the solid, dash-dotted, dashed and dotted curves respectively.}\label{TD_funct}
\end{figure}

In this section we consider thermodynamics of the Dirac oscillator. It should be pointed out that thermal properties of Dirac oscillators in various dimensions were examined in literature \cite{Boumali_arx15,Pacheco_EPL14,Frassino_arx17} earlier. Similarly to the mentioned above papers we consider the Dirac oscillator in a thermal bath so to describe the thermal properties we use canonical partition function which can be written in the form:
\begin{equation}\label{part_funct}
Z=\sum^{N_{max}}_{n=0}e^{-\frac{E_n}{T}}
\end{equation}
and here $E_n$ is the energy spectrum given by the relations (\ref{energy_z}) or (\ref{en_sp_2}) and $N_{max}$ is the upper value of the quantum number $n$ which is restricted by the Planck energy (momentum). In the following we consider the first branch of the spectrum (\ref{energy_z}) because its well-defined nondeformed limit, the thermal properties for the second branch of the spectrum will be investigated elsewhere. Due to complicated form of the relation (\ref{energy_z}) it is not possible to calculate the partition function (\ref{part_funct}) exactly, so some approximation should be used. Firstly, we simplify the expression for the spectrum (\ref{energy_z}) taking into account the first nonzero corrections caused by the deformation of the commutation relations. This simplified form of the energy spectrum (\ref{energy_z}) can be represented as follows:
\begin{equation}\label{sp_simp}
E_n\simeq m\sqrt{an^2+bn+1},
\end{equation}
where we denoted $a=7\o^2\a^2/4$ and $b=2\o/m$. It should be pointed out here that when one takes into account the minimal length only the latter relation (\ref{sp_simp}) is exact and $a=2\o^2\a^2$ (it can be seen from the relation (\ref{energy_z})). The simplified form of the spectrum (\ref{sp_simp}) is still complicated enough to allow the partition function to be calculated exactly. To calculate it we make use of Euler-Maclaurin summation relation:
\begin{equation}
\sum^{N}_{j=n}f(j)=\int^{N}_{n}f(x)dx+\frac{1}{2}(f(n)+f(N))+\sum^{+\infty}_{j=2}\frac{B_j}{j!}\left(f^{(j-1)}(N)-f^{(j-1)}(n)\right),
\end{equation}
where $B_j$ are Bernoulli numbers. As a result the partition function can be represented in the form:
\begin{eqnarray}\label{Z_dec}
\nonumber Z=\int^{N_{max}}_{0}e^{-\b\sqrt{ax^2+bx+1}}dx+\frac{1}{2}\left(1+e^{-\b\sqrt{aN^2_{max}+bN_{max}+1}}\right)+\\\sum^{+\infty}_{j=2}\frac{B_j}{j!}\left(\frac{d^{j-1}}{dx^{j-1}}e^{-\b\sqrt{ax^2+bx+1}}\Big|_{x=N_{max}}-\frac{d^{j-1}}{dx^{j-1}}e^{-\b\sqrt{ax^2+bx+1}}\Big|_{x=0}\right)
\end{eqnarray}
and here $\b=m/T$. Integral in the latter relation can not be calculated exactly. To obtain analytical expression we suppose that the deformation parameter is small and corresponding term in the spectrum is substantially smaller than nondeformed part. So this integral can be rewritten in the form:
\begin{equation}
\int^{N_{max}}_{0}e^{-\b\sqrt{ax^2+bx+1}}dx\simeq \int^{N_{max}}_{0}e^{-\b\sqrt{bx+1}}\left(1-\frac{\b a x^2}{2\sqrt{1+bx}}\right)dx
\end{equation} 
When the $N_{max}$ is large enough but finite the contribution of the upper bound in the written above integral is negligibly small and for simplicity we will not take it into account. So, for the written above integral we have:
\begin{equation}
\int^{N_{max}}_{0}e^{-\b\sqrt{bx+1}}\left(1-\frac{\b a x^2}{2\sqrt{1+bx}}\right)dx\simeq\frac{2}{\b b}\left(1+\frac{1}{\b}-\frac{4a}{\b b^2}\left(1+\frac{3}{\b}+\frac{3}{\b^2}\right)\right)e^{-\b}
\end{equation} 
Similarly we can neglect by the contribution from the upper bound in the approximated relation for the partition function (\ref{Z_dec}). Taking into account only first two terms in the sum of the Euler-Maclaurin relation we can represent the partition function in the form:
\begin{eqnarray}\label{Z_final}
Z=\frac{1}{2}+\frac{2}{\b b}\left(1+\frac{1}{\b}-\frac{4a}{\b b^2}\left(1+\frac{3}{\b}+\frac{3}{\b^2}\right)\right)e^{-\b}+\frac{\b b}{24}e^{-\b}+\frac{\b b}{1440}e^{-\b}\left(3(1+\b)\left(a-\frac{b^2}{4}\right)-\frac{\b^2 b^2}{4}\right).
\end{eqnarray}
We point out here that similar relation we obtain in case of a deformed algebra with minimal length only, but the only difference between that and our present form of partition function is hidden in different values of the parameter $a$. Having calculated the partition function and using standard relations of statistical physics one can obtain thermodynamic functions of the Dirac oscillator. Namely we consider the following functions:
\begin{eqnarray}
F=-T\ln{Z}, \quad U=-T^2\frac{\partial}{\partial T}\left(\frac{F}{T}\right),\\ S=-\left(\frac{\partial F}{\partial T}\right)_V, \quad C_V=T\left(\frac{\partial S}{\partial T}\right)_V.
\end{eqnarray}
Using the written above relations one can derive analytical expressions  for the corresponding functions but due to the complicated relation for the partition function (\ref{Z_final}) we perform it numerically.  The Fig.[\ref{TD_funct}] shows all these thermodynamic values as functions of dimensionless temperature $T/m$. One can see that the behaviour of thermodynamic functions considerably depend on the value of the parameter of deformation $\a$, namely small values of the parameter $\a$  allow to recover the behaviour of the thermodynamic functions for nondeformed Dirac oscillator whereas large values of $\a$ might make the system thermodynamically unstable, or stable in quite narrow range of temperatures. Qualitatively the behaviour of the thermodynamic functions is similar to the behaviour of the related functions in case of $q$-deformed commutation relations or deformed algebra with minimal length \cite{Boumali_AHEP17}.
\section{Conclusions}
We have considered $1+1$-dimensional Dirac oscillator with deformed algebra with minimal uncertainty in position and maximal uncertainty in momentum. The minimal uncertainty in position as is known leads to the consequence that the corresponding deformed algebra possesses the momentum representation only \cite{Kempf_PRD95}, whereas the deformed algebra with minimal uncertainty in position and maximal in momentum gives rise to the conclusion that we have the upper bound for momentum \cite{Ali_PLB09,Nozari_PRD12} and as a result the upper bound for the energy. The upper bound for the energy gives rise to the consequence that energy spectrum might have a finite numbers of levels, what we have in our case. To find the energy spectrum of the Dirac oscillator we have used the SUSY QM technique which allows to construct iterative relation for  the gaps between the energy levels $\ve_i$ and obtain the energy spectrum in a simple way. The SUSY QM technique also allows to derive eigenfunctions of the system but we have restricted oneself by the ground state wavefunctions and shown that they would have well-defined norm.  We have shown that the energy spectrum of the Dirac oscillator in deformed case has two branches, namely the first one is the so-called branch with zero ground state energy (\ref{energy_z}) and in the limit when $\a\rightarrow 0$ we recover the spectrum of the standard nondeformed Dirac oscillator. The second branch is the branch with nonzero ground state energy (\ref{en_sp_2}), but this branch does not have the limit $\a\rightarrow 0$.

In  the second part of  our work we have examined some thermal properties of the Dirac oscillator. Similarly as it was done earlier we have considered the Dirac oscillator in the canonical ensemble framework. Because of complicated form of the partition function we have taken into consideration only the first order corrections over the deformation parameter. Here we consider only the first branch of the spectrum, the thermal properties of the oscillator with the second branch of the spectrum will be considered elsewhere. Having calculated the partition function we have performed numerical calculation of the main thermodynamic functions, namely of the free energy, internal energy, entropy and heat capacity for several different values of the parameter of deformation $\a$. It has been shown that the mentioned above functions in case of a small parameter of deformation can approximate corresponding nondeformed values, whereas for large parameter of deformation the given functions have substantially distinct behaviour  which shows that the system might be thermodynamically unstable or exist in some range of temperatures.

\section{Acknowledgements}
This work was partly supported by Project FF-30F (No. 0116U001539) from the Ministry of Education and Science of Ukraine and Project STREVCOMS PIRSES-2013-612669 from the European Commission.

\end{document}